\documentclass[aps,prc,preprint,superscriptaddress,nofootinbib]{revtex4-2}
\usepackage{graphicx}
\usepackage{dcolumn}
\usepackage{bm}
\usepackage{epstopdf}
\usepackage{amsmath}
\usepackage{hyperref}
\usepackage{color}
\usepackage{cancel}

\begin{document}

\title{The 25th anniversary for nuclear chirality}

\author{J. Meng}
\affiliation{State Key Laboratory of Nuclear Physics and Technology, School of Physics, Peking University, Beijing 100871, China}
\author{Y. P. Wang}
\affiliation{State Key Laboratory of Nuclear Physics and Technology, School of Physics, Peking University, Beijing 100871, China}

\date{\today}

\maketitle

\date{today}

\section{Introduction}

Chirality is a subject of general interests in natural science.
Since the concept of the chirality in atomic nuclei  \cite{Frauendorf1997NPA} was proposed 25 years ago, it has become one of the hot topics in modern nuclear physics. 
The history of the prediction of the nuclear chirality is quite instructive.

The gestation of the concept of the nuclear chirality can be dated back to about 30 years ago when one of the authors J. M. visited Forschungszentrum Rossendorf, now called Helmholtz Zentrum Dresden Rossendorf, in Germany. J. M. stayed there from September 1993 to August 1994. J. M. enjoyed the time with S. Frauendorf and other colleagues very much, and they published five papers. 

\begin{figure*}[h]
  \centering
  \includegraphics[width=0.9\linewidth]{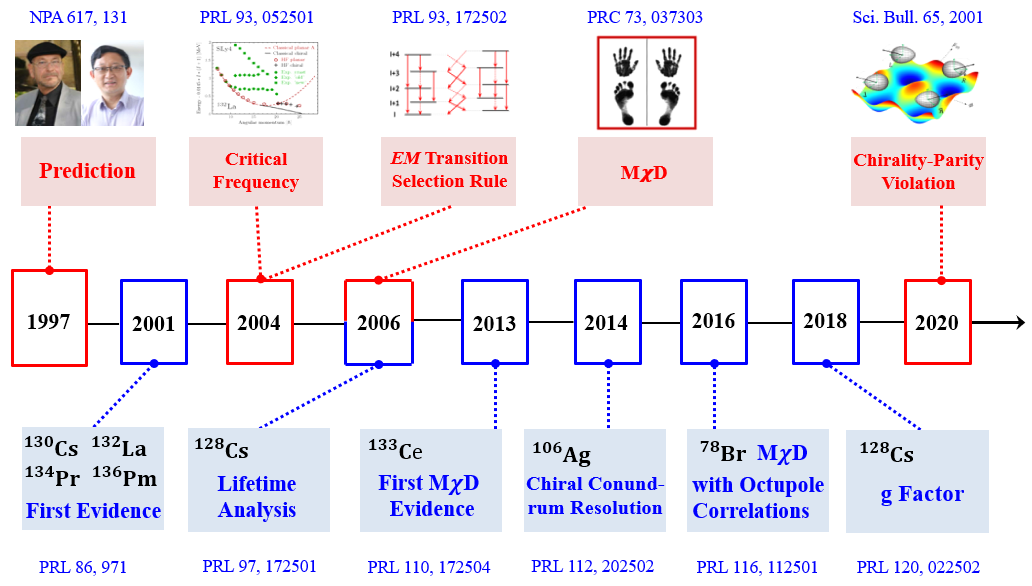}
  \caption{The time line for some of the significant theoretical (top) and experimental (bottom) progresses in nuclear chirality.}
  \label{fig11}
\end{figure*}

At that time, one of the hot research topics was the superdeformed rotational band \cite{Twin1986PRL}, while the investigation of the magnetic rotation was only at its beginning. 
Traditionally, it is well-known that the rotations occur only in nuclei with a stable deformation. Thus the observation of the regular rotational-like sequences in several near spherical Pb isotopes was very surprising (for a review see Ref. \cite{Hubel2005PPNP}). 
The explanation of such bands was first given by S. Frauendorf using the axial tilted axis cranking (TAC) approach in 1993 \cite{Frauendorf1993NPA}. 
It was found that the angular momentum along these bands is generated by the alignment of the proton and neutron angular momenta. 
Unlike the normal deformed rotational bands, these rotational-like bands have strong magnetic dipole ($M1$) and very weak electric quadrupole ($E2$) transitions, thus the name ``magnetic rotation'' was introduced \cite{Frauendorf1994Conference}. 

After the concept of magnetic rotation was proposed, many efforts have been devoted to verify it experimentally.  
However, there are big discrepancies between the predicted and observed angular momentum dependence of the $B(M1)$ values \cite{Neffgen1995NPA}. 
Therefore, at that time, the results of the TAC model were questioned, due to its semi-classical treatment of the angular momentum, the mean-field approximation, and the neglect of the multiparticle correlations. 

After arriving in Rossendorf for one month, J. M. examined the quality of the semi-classical and mean-field approximation of the TAC model.
With the recommendation of S. Frauendorf, J. M. attended the European Centre for Theoretical Studies in Nuclear Physics and Related Areas (ECT*) workshop in 1993 and presented the results, which was later published in Zeitschrift f{\"u}r Physik A titled ``Interpretation and quality of the tilted axis cranking approximation'' in 1996 \cite{Frauendorf1996ZPA} . In this paper, the quality of the semi-classical and mean-field approximation of the TAC model was examined by comparing with the results from the particle rotor model (PRM). 

The PRM is a quantum model with the total angular momentum as a good quantum number, works in the laboratory frame, and describes properly the  quantum tunneling. Surprisingly, the calculated energy spectrum as well as the $B(M1)$ and $B(E2)$ values in PRM are well reproduced by the TAC model \cite{Frauendorf1996ZPA}. 
Thus the semi-classical and mean-field approximation are not responsible for the deviation between the calculated and the observed $B(M1)$ values. 
Later on, with the improvement of the detector technology, the predicted decrease of the $B(M1)$ values with the angular momentum was turned out to be correct \cite{Clark1997PRL}. 

Following the success of the TAC model for the magnetic rotation, it is quite natural to generalize it for the triaxial nuclei. The famous paper, in which the concept of the nuclear chirality was suggested, appeared by using the triaxial TAC model and the PRM \cite{Frauendorf1997NPA}.

Since then, the nuclear chirality has become one of the hot topics in modern nuclear physics. 
Over the past two and a half decades, many efforts have been devoted to understand the nuclear chirality, and some of the significant theoretical and experimental progresses are summarized in a time line in Fig. \ref{fig11}.  

In this chapter, the prediction of the nuclear chirality is presented in Sec.~\ref{sec2}.
The theoretical and experimental investigations of the nuclear chirality are reviewed, including the verification of chiral doublet bands in Sec.~\ref{sec3}, the chiral conundrum and its resolution in Sec.~\ref{sec4}, and the prediction and observation of the multiple chiral doublets (M$\chi$D) in Secs.~\ref{sec5}-\ref{sec6}.
Some recent theoretical progresses are highlighted in Sec.~\ref{sec7} , including the chiral collective Hamiltonian, the $A$-plot and the $K$-plot, the nuclear chirality-parity (ChP) violation, the chiral rotation induced by the pairing correlations, as well as the chiral dynamics.
The possibly emerging area, challenges that lie ahead, and opportunities for progress in the context of the nuclear chirality are discussed in Sec.~\ref{sec8}.

\section{Prediction of Nuclear Chirality}
\label{sec2}

\begin{figure*}[!]
  \centering
  \includegraphics[width=0.4\linewidth]{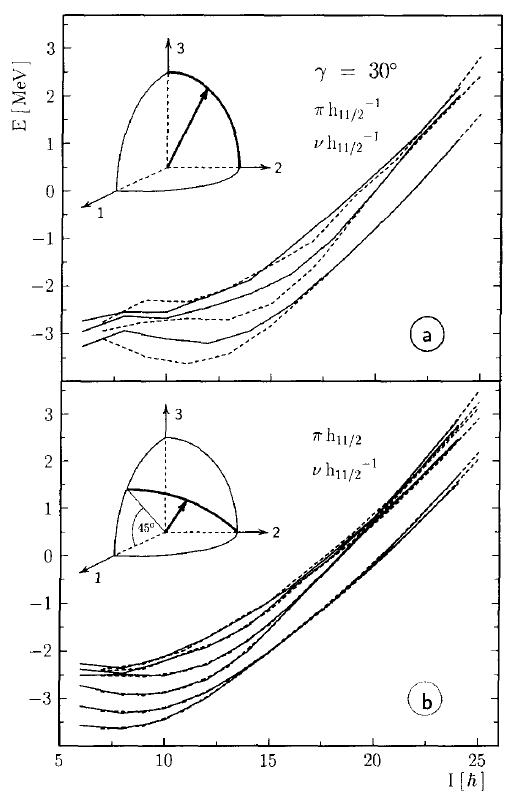}
  \caption{ Rotational levels of $h_{11/2}$ particles and holes coupled to a triaxial rotor with $\gamma = 30^{\circ}$. The upper panel shows the case of a proton hole and a neutron hole and the lower panel shows the case of a proton particle and a neutron hole. Full lines correspond to even and dashed to odd spins. Reprinted figure from \cite{Frauendorf1997NPA}.}
  \label{fig3}
\end{figure*}

In Ref. \cite{Frauendorf1997NPA}, the nucleus is assumed to have a high-$j$ particle and a high-$j$ hole coupled with a triaxial rotor. The angular momenta for the particle and the hole respectively favor to align along the nuclear short ($s$) and long ($l$) axes. The angular momentum for the rotor favors to align along the intermediate ($i$) axis due to its largest moment of inertia. 
In the TAC model, the total angular momentum vector may lie outside the three principal planes of the triaxial ellipsoidal density distribution. The $s$, $i$ and $l$ principal axes of the triaxial nucleus form a screw with respect to the angular momentum vector, resulting in two systems with different intrinsic chirality, i.e., left- and right-handed systems. In the PRM, the broken chiral symmetry in the intrinsic frame would be restored, and this gives rise to the chiral doublet bands, i.e., a pair of nearly degenerate $\Delta I=1$ bands with the same parity, as shown in Fig. \ref{fig3}. 

Before the publication of Ref. \cite{Frauendorf1997NPA}, the examples for the pair of nearly degenerate bands had already been reported in $^{134}$Pr \cite{Petrache1996NPA}. In a conference organized by Rainer Lieder and others in the beautiful Crete island in Greece, after the first session by the talk of Frauendorf and J. M., Petrache appeared and showed his results. Therefore, the observed partner bands in $^{134}$Pr were suggested as a candidate for chiral doublet bands in Ref. \cite{Frauendorf1997NPA}.

\begin{figure*}[!]
  \centering
  \includegraphics[width=0.9\linewidth]{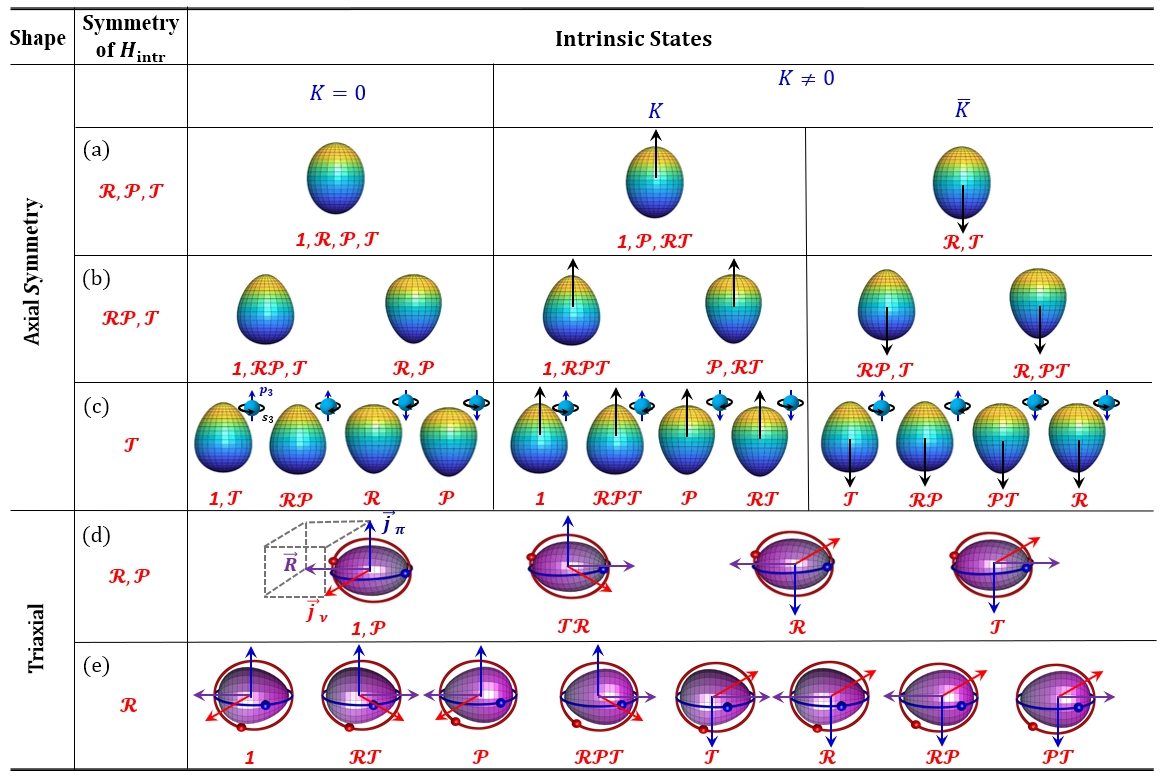}
  \caption{The intrinsic states for axially symmetric and triaxial systems with different combinations of $\mathcal{P}$, $\mathcal{T}$ and $\mathcal{R}$ invariance. Adapted figure from Ref. \cite{Bohr1975NuclearII}.}
  \label{fig1}
\end{figure*}

\begin{figure*}[!]
  \centering
  \includegraphics[width=0.9\linewidth]{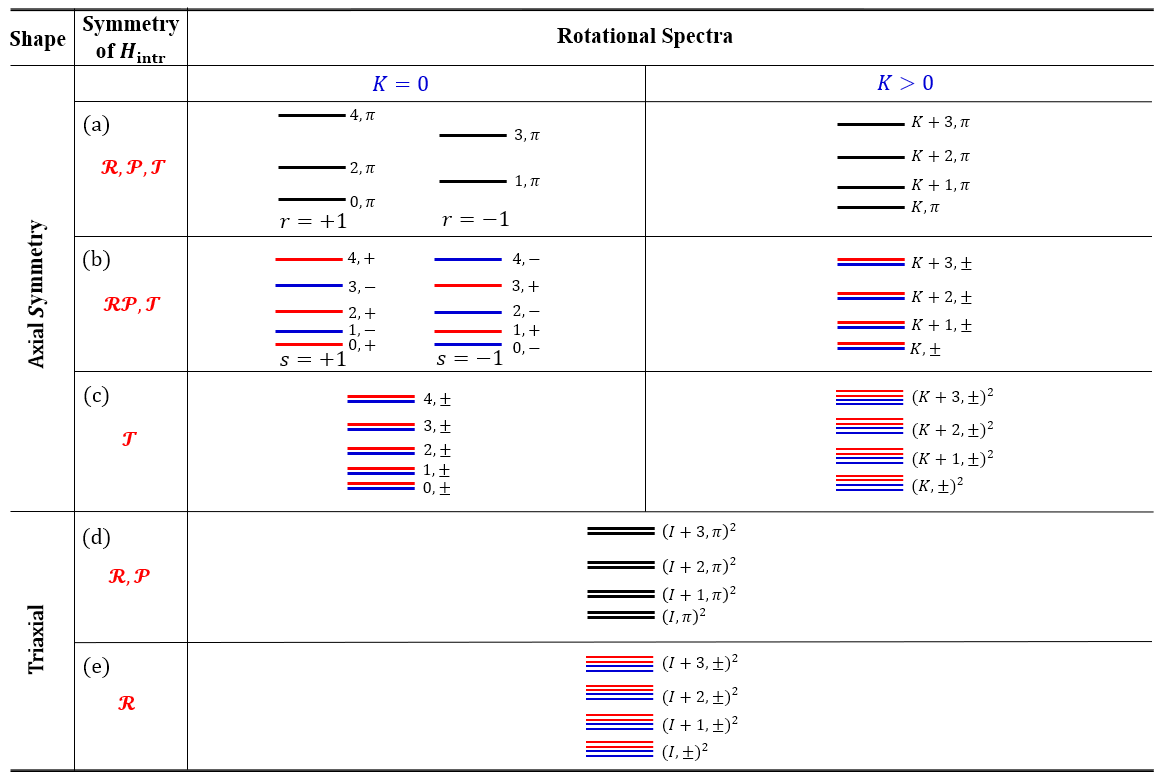}
  \caption{The rotational spectra for axially symmetric and triaxial systems with different combinations of $\mathcal{P}$, $\mathcal{T}$ and $\mathcal{R}$ invariance. The quantum numbers $(I,\pi)$ are given to the right of the energy levels, with $I$ the angular momentum and $\pi$ the parity. Adapted figure from Ref. \cite{Bohr1975NuclearII}.}
  \label{fig2}
\end{figure*}

In the textbook \textit{Nuclear Structure} by Bohr and Mottelson, the structure of the rotational spectra is determined by the symmetry of the nuclear deformation and the resulting rotational degrees of freedom \cite{Bohr1975NuclearII}. 
The nuclear system with quadrupole deformation may possess the parity ($\mathcal{P}$), time-reversal ($\mathcal{T}$), spatial rotation of $180^{\circ}$ ($\mathcal{R}$) symmetries and their combinations. 
The cases when the system possesses the symmetries $\mathcal{R}, \mathcal{P}, \mathcal{T}$, the symmetries $\mathcal{RP}, \mathcal{T}$, or only the symmetry $\mathcal{T}$ have been discussed in Ref. \cite{Bohr1975NuclearII}, as shown in Figs. \ref{fig1} (a-c).
The associated rotational spectra are schematically shown in Figs. \ref{fig2} (a-c).
The system with the symmetries $\mathcal{R}$ and $\mathcal{P}$ hasn't been discussed in Ref. \cite{Bohr1975NuclearII}, which corresponds to the nuclear chirality, and the corresponding symmetry as well as the associated rotational spectra are respectively shown in Fig. \ref{fig1} (d) and Fig. \ref{fig2} (d).

\section{Observation of Nuclear Chirality}
\label{sec3}

After the prediction in 1997, lots of experimental efforts
have been devoted to search for the nuclear chirality. 
In 2001, four pairs of nearly degenerate bands were observed in the odd-odd $N=75$ isotones $^{130}$Cs, $^{132}$La, $^{134}$Pr and $^{136}$Pm respectively, and were suggested as the candidate chiral doublet bands \cite{Starosta2001PRL}. 
Two years later, similar chiral doublet bands were observed in odd-$A$ nucleus \cite{ZhuS2003PRL}. 
Based on the observed near-degenerate $\Delta I = 1$ doublet bands with the same parity, candidate chiral doublet bands have been proposed in a number of nuclei in the $A =$ 130 region with the configuration $\pi h_{11/2}\otimes \nu h_{11/2}^{-1}$ \cite{Petrache1996NPA,Koike2001PRC,Hecht2001PRC,Hartley2001PRC,Bark2001NPA,Mergel2002EPJA,Starosta2002PRC,LiXF2002CPL,ZhuS2003PRL,Koike2003PRC,Rainovski2003PRC,Simons2005JPG,WangSY2006PRC}, $A =$ 100 region with $\pi g_{9/2}^{-1}\otimes \nu h_{11/2}$ \cite{Vaman2004PRL,Alcantara-Nunez2004PRC,Timar2004PLB,Joshi2005EPJA,ZhuS2005EPJA,Timar2006PRC,HeCY2006HEPNP,Joshi2007PRL,Timar2007PRC}, $A =$ 190 region with $\pi h_{9/2}\otimes \nu i_{13/2}^{-1}$ \cite{Balabanski2004PRC,Lawrie2008PRC}. 
In 2011, under joint efforts from scientists in China, South Africa and Hungary, the first example of chiral doublet bands based on the configuration $\pi g_{9/2}\otimes \nu g_{9/2}^{-1}$ was reported in the $A =$ 80 region \cite{WangSY2011PLB}, which is the lightest known region with chiral nuclei. 

In the same period, the significant progress has also been made in the study of the nuclear chirality on the theoretical side. 
A self-consistent three-dimensional (3D) mean-field solution has been obtained and strongly supports the possible existence of the chiral rotation in actual nuclei \cite{Dimitrov2000PRL}. 
The selection rule for the electromagnetic transitions expected in the ideal chiral bands has been proposed, including the odd-even staggering of intraband $B(M1)/B(E2)$ ratios and interband $B(M1)$ values, as well as the vanishing of the interband $B(E2)$ transitions at high spin region \cite{Koike2004PRL}. 
The critical frequency which marks the onset of chiral rotation was suggested by the Skyrme Hartree-Fock cranking model calculations \cite{Olbratowski2004PRL}. 

\section{Chiral Conundrum in $^{134}$Pr}
\label{sec4}

With more and more chiral doublet candidates observed, it is natural to measure other observables than the energy spectra to provide further confirmation for the chirality. 
In 2006, the electromagnetic transition ratios of the nearly degenerate bands in $^{134}$Pr were measured \cite{Tonev2006PRL}. 
The measured in-band $B(E2)$ values for the candidate partner bands show large differences, and this is in disagreement with the chiral picture. This phenomenon was known as chiral conundrum.
In particular, the paper titled with ``Risk of misinterpretation of nearly degenerate pair bands as chiral partners in Nuclei'' \cite{Petrache2006PRL} casted a shadow on the investigation of the nuclear chirality.

It should be noted that although the chiral partner bands have energies close to each other, it is rare to observe a crossing between them. The most famous example of such a crossing is in $^{134}$Pr. The only other known case is in $^{106}$Ag \cite{Joshi2007PRL}. Thus the problem in $^{134}$Pr might be a special case, and may not influence the investigation of the nuclear chirality. 

Fortunately, shortly after that, lifetime measurement was performed for $^{128}$Cs and the electromagnetic transition properties of the partner bands in $^{128}$Cs were in good agreement with the chiral interpretation \cite{Grodner2006PRL}. 
It was claimed as the best known example revealing the chiral symmetry breaking phenomenon. 
Later on, the chiral character of the bands in $^{135}$Nd was also affirmed based through the lifetime measurements and it was shown that the partner bands are associated with a transition from chiral vibration to static chirality \cite{Mukhopadhyay2007PRL}. 
These observations renewed the enthusiasm on the study of the nuclear chirality.

\section{Prediction of M$\chi$D}
\label{sec5}

It should be emphasized that caution should be taken when comparing the theoretical results with the experimental data.
The nuclear chirality was predicted in an ideal model, i.e., a high-$j$ particle and a high-$j$ hole coupled with a triaxial rotor with $\gamma=30^{\circ}$. 
However, it is not specified which nuclei satisfy these conditions. 
In order to search for the chiral nucleus, the microscopic approaches with predictive power are demanded. 

Starting from an effective nucleon-nucleon interaction with Lorentz invariance, the covariant density functional theory (CDFT) is very successful in describing many nuclear phenomena in stable and exotic nuclei of the whole nuclear
chart \cite{Ring1996PPNP,Vretenar2005PR,Niksic2011PPNP,MengJ2016Relativistic}. 
It provides a powerful way for the investigation of the nuclear chirality. 

\begin{figure*}[!]
  \centering
  \includegraphics[width=0.9\linewidth]{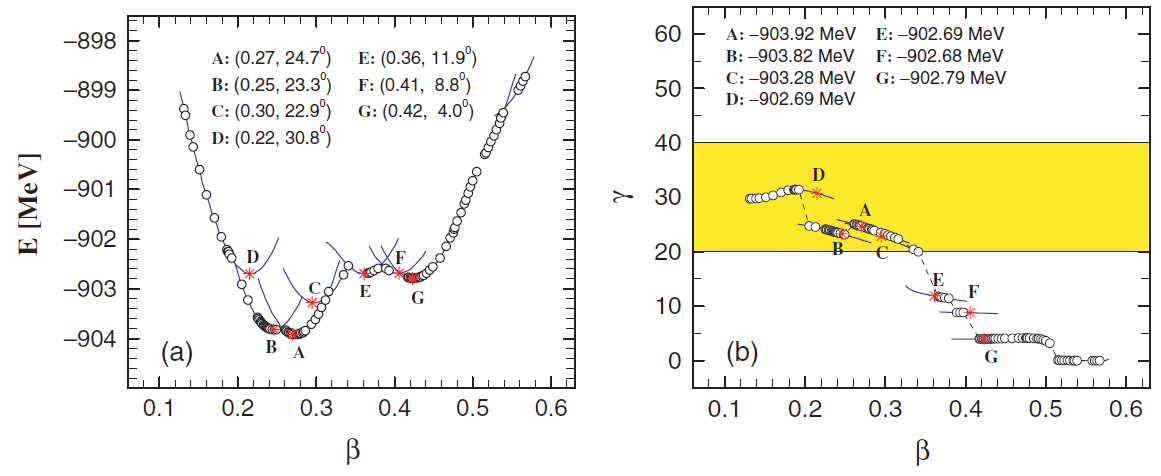}
  \caption{Energy surfaces (a) and deformation parameters $\gamma$ (b) as functions of deformation parameter $\beta$ in 
           adiabatic (open circles) and configuration-fixed (solid lines) constrained triaxial CDFT calculations for $^{106}$Rh. The minima in the energy surfaces are represented as stars and labeled A-G with their corresponding deformations $\beta$ and $\gamma$ (a) and energies (b). Reprinted figure from \cite{MengJ2006PRC}.}
  \label{fig4}
\end{figure*}

In 2006, the adiabatic and configuration-fixed constrained triaxial CDFT was developed to predict chiral nuclei and guide the phenomenological models \cite{MengJ2006PRC}. The nucleus $^{106}$Rh was investigated as an example. 
The energy surfaces and deformation $\gamma$ for $^{106}$Rh are shown as functions of deformation $\beta$ in Fig. \ref{fig4}. 
It was found that the minima, A, B, C, and D, have deformations $\beta$ and $\gamma$ suitable for chirality.
Their proton and neutron configurations were also examined in detail to see whether they correspond to the high-$j$ particle and hole configurations required by chirality.
Then a new phenomenon, which was called as the multiple chiral doublets (M$\chi$D) , i.e., more than one pair of chiral doublet bands in one single nucleus, was suggested for $^{106}$Rh.

\section{Observation of M$\chi$D}
\label{sec6}

The prediction of M$\chi$D stimulated lots of experimental efforts. 
In 2013, two distinct sets of chiral partner bands were identified in $^{133}$Ce, which was the first experimental evidence for M$\chi$D \cite{Ayangeakaa2013PRL}. 
Along this line, the lifetime measurements for the chiral doublet bands in $^{106}$Ag provided a resolution of the chiral conundrum \cite{Lieder2014PRL}. 
Similarly, the chirality in $^{134}$Pr needs further investigation.
In 2014, the M$\chi$D with the identical configuration were observed in $^{103}$Rh \cite{Kuti2014PRL}, showing that the chiral geometry in nuclei can be robust against the increase of the intrinsic excitation energy. 
In 2016, the M$\chi$D with octupole correlations were first identified in $^{78}$Br \cite{LiuC2016PRL}. 
This observation indicates that the nuclear chirality can be robust against the octupole correlations. 
It also indicates that the chirality-parity (ChP) quartet bands \cite{Frauendorf2001RMP}, which are a consequence of the simultaneous breaking of chiral and space-reflection symmetries, namely, ChP violation, may exist in nuclei.

\begin{figure*}[!]
  \centering
  \includegraphics[width=0.9\linewidth]{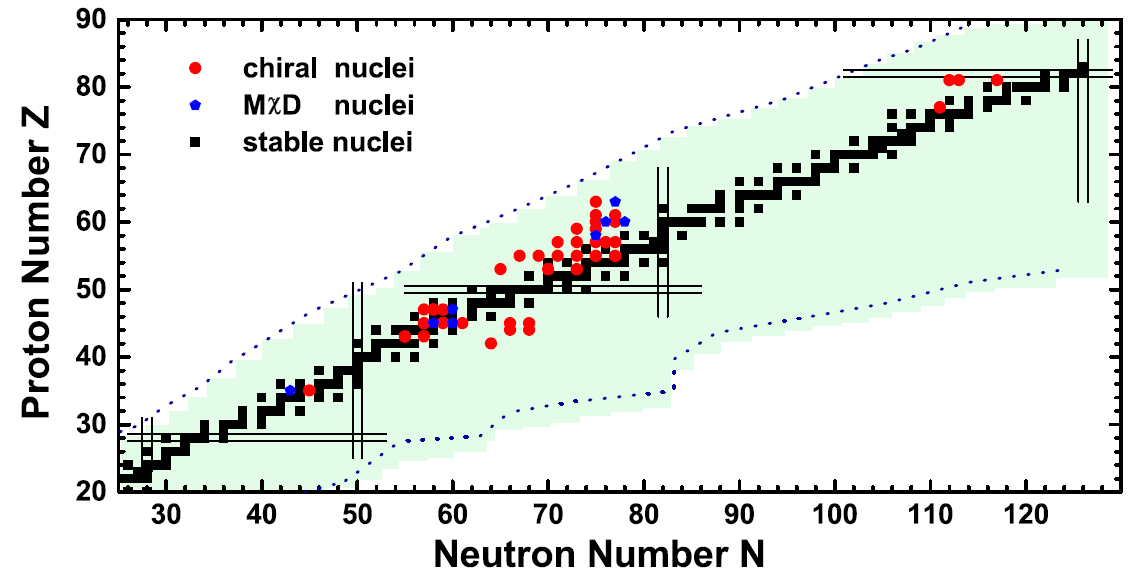}
  \caption{The nuclides with chiral doublet bands (circles) and M$\chi$D (pentagons) observed in the nuclear chart. 
         The squares represent stable nuclides. Reprinted figure from \cite{XiongBW2019ADNDT}.}
  \label{fig5}
\end{figure*}

Until 2019, more than 59 chiral doublet bands in 47 nuclei (including 8 nuclei with M$\chi$D) have been reported in the $A=$ 80, 100, 130, and 190  regions \cite{XiongBW2019ADNDT}. 
The distribution of the observed chiral nuclei in the nuclear chart is given in Fig. \ref{fig5}.  

\section{Recent Progress}
\label{sec7}

The nuclear chirality has been extensively investigated with many theoretical approaches, including the triaxial PRM \cite{Frauendorf1997NPA,PengJ2003PRC,Koike2004PRL,ZhangSQ2007PRC,QiB2009PLB,ChenQB2018PLB,WangYY2019PLB,WangYP2020PRC,WangYY2020SB},
the 3D cranking model \cite{Frauendorf1997NPA,Madokoro2000PRC,Dimitrov2000PRL,Olbratowski2004PRL,ZhaoPW2017PLB,PengJ2020PLB,WangYP2023PLB,RenZX2022PRC}, the 3D cranking model with the random phase approximation \cite{Mukhopadhyay2007PRL,Almehed2011PRC}, the 3D cranking model with the collective Hamiltonian \cite{ChenQB2013PRC,ChenQB2016PRC,ChenQB2018PRC}, the interacting boson-fermion-fermion model  \cite{Brant2008PRC}, the generalized coherent state model \cite{Raduta2016JPG},
and the projected shell model \cite{Hara1995IJMPE,Bhat2014NPA,ChenFQ2017PRC,ChenFQ2018PLB,WangYK2019PRC}.
Some of these progresses are highlighted in this Section.

\subsection{Chiral Collective Hamiltonian}\label{collective}

Within the TAC mean-field approximation, the left-handed and right-handed solutions are exactly degenerate. 
It is not possible to calculate the energy difference between the bands due to the missing of the quantum tunneling. 

To describe the energy splitting between the chiral doublet bands, a collective Hamiltonian for chiral modes was constructed in analogy to Bohr Hamiltonian \cite{ChenQB2013PRC}. 
Instead of the $\beta$ and $\gamma$ degrees of freedom in Bohr Hamiltonian, the azimuthal angle $\phi$ was introduced as the collective degree of freedom. 
The TAC with the collective Hamiltonian model reproduced well the energy spectra for the chiral partners obtained by the PRM. 
By considering both the azimuthal angle $\phi$ and the polar angle $\theta$ as the collective degrees of freedom, the collective Hamiltonian was extended to two-dimensional, and more excitation modes have been obtained \cite{ChenQB2016PRC,WuXH2018PRC}.  

In the future, the collective Hamiltonian model could be combined with the microscopic 3D cranking CDFT.  The collective Hamiltonian on top of the 3D cranking CDFT could become very powerful in describing and predicting chiral doublet bands.

\subsection{$A$-plot and $K$-plot}\label{KA}

The projected shell model (PSM) restores the rotational symmetry broken in the mean-field approximation and combines the advantages of the TAC model and the PRM. 

The attempts to understand the chiral doublet bands by the PSM have been performed in Refs. \cite{Bhat2012PLB,Bhat2014NPA} where the energy spectra and transitions are well reproduced. However, it is a big challenge to examine the chiral geometry of angular momentum in the PSM due to the complication that the projected basis is defined in the laboratory frame and forms a nonorthogonal set.

In Refs. \cite{ChenFQ2017PRC,ChenFQ2018PLB}, the orientation of the angular momentum in the intrinsic frame is investigated by the distributions of its components on the three principal axes, $K$-plot, and those of its tilted angles, azimuthal plot ($A$-plot). The evolution of the chirality with spin is illustrated, and the chiral geometry is demonstrated in the PSM.

\begin{figure*}[!]
  \centering
  \includegraphics[width=0.9\linewidth]{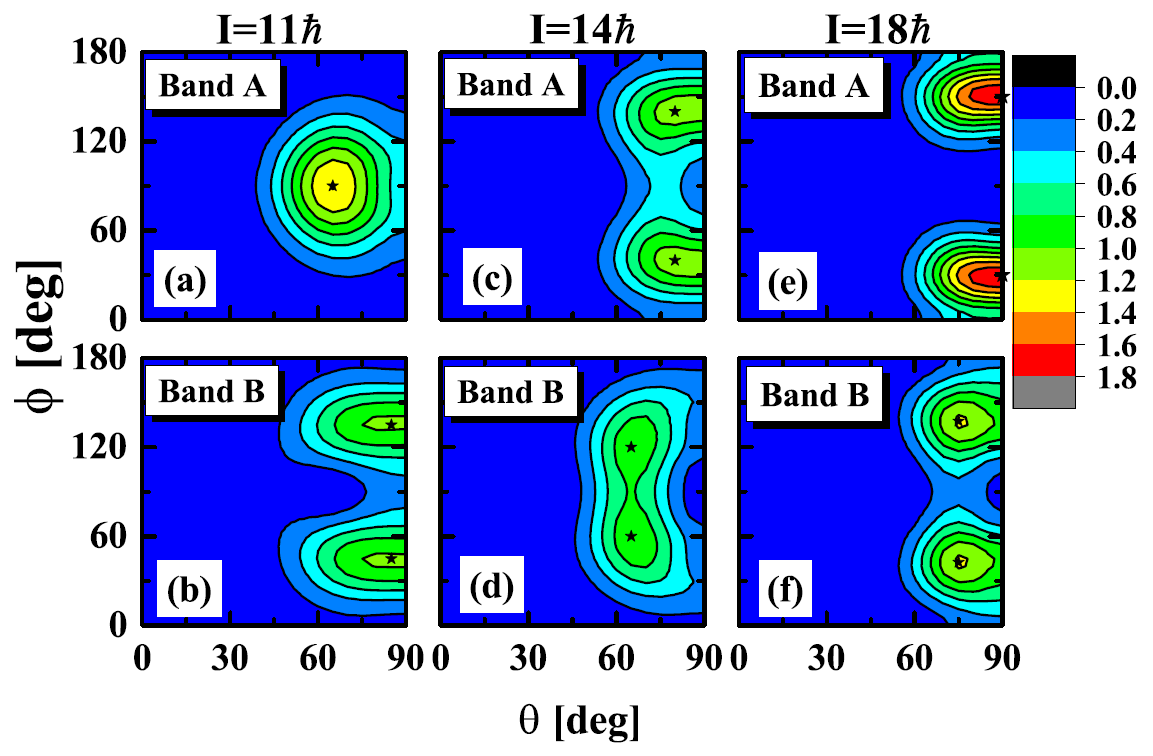}
  \caption{The azimuthal plots ($A$-plots), i.e., profile for the orientation of the angular momentum on the $(\theta,\phi)$ 
           plane, calculated at $I=11$ [(a), (b)], $14$ [(c), (d)], and $18\hbar$ [(e), (f)], respectively. Reprinted figure from \cite{ChenFQ2017PRC}.}
  \label{fig6}
\end{figure*}

The $A$-plots for the chiral doublet bands A and B in $^{128}$Cs are shown in Fig. \ref{fig6}. 
These plots clearly illustrate the evolution of the chiral mode from a chiral vibration at $I=11\hbar$ with respect to the $l$-$s$ plane, to a static chiral rotation at $I=14\hbar$, and to another chiral vibration with respect to the $i$-$s$ plane at $I=18\hbar$.  

\subsection{Nuclear ChP Violation}\label{CP-Violation}

The observation in $^{78}$Br, two pairs of chiral doublet bands with opposite parity connected with strong electric dipole ($E1$) transitions,
provides the first evidence of M$\chi$D in octupole soft nuclei \cite{LiuC2016PRL}. This observation indicates that the nuclear chirality can be robust against the octupole correlations, which together with the scenario in Ref. \cite{Frauendorf2001RMP} encourages the exploration of the simultaneous chiral and reflection symmetry breaking, ChP violation, in a reflection asymmetric triaxial nucleus.

\begin{figure*}[!]
  \centering
  \includegraphics[width=0.6\linewidth]{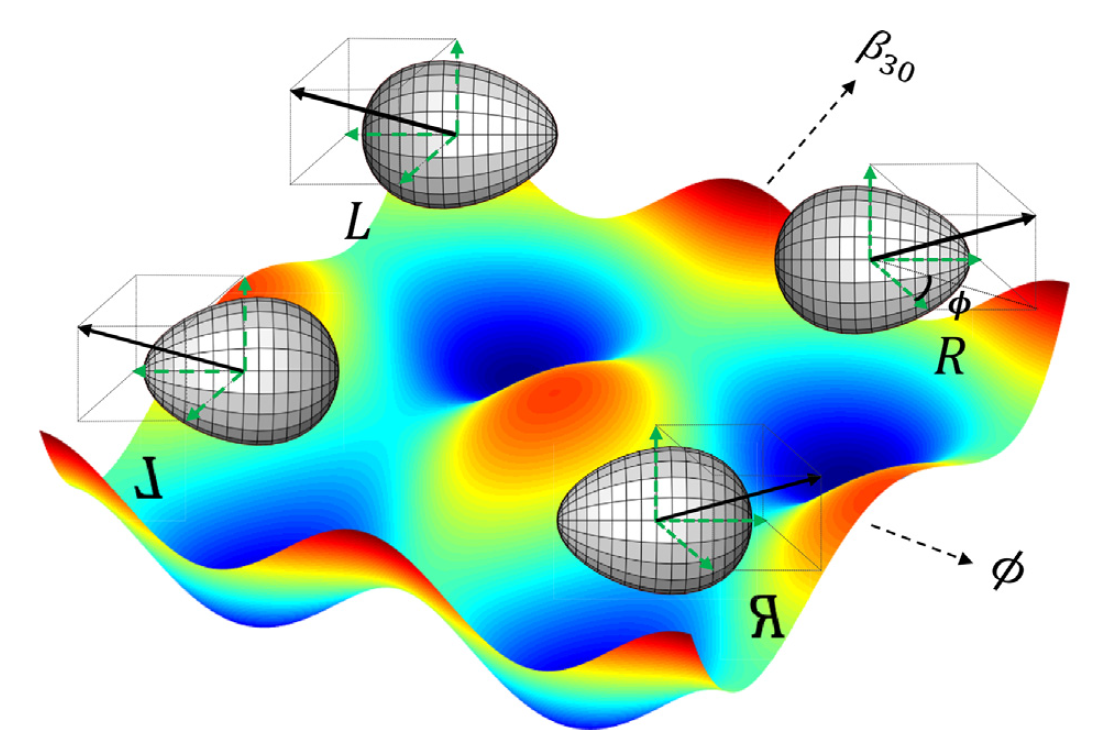}
  \caption{A schematic potential energy surface in the intrinsic $(\beta_{30},\phi)$ plane for a ChP violation system, with 
           $\beta_{30}$ the octupole deformation parameter and $\phi$ the azimuthal angle of the total angular momentum. The sign of $\beta_{30}$ stands for the orientation of the nuclear distribution parallel or antiparallel with the intrinsic axis. The sign of $\phi$ stands for the right- and left-handed system. Reprinted figure from \cite{WangYY2020SB}.}
  \label{fig7}
\end{figure*}

A schematic potential energy surface with simultaneous chiral and reflection symmetry breaking in the intrinsic $(\beta_{30},\phi)$ plane is given in Fig. \ref{fig7}, with $\beta_{30}$ the octupole deformation parameter and $\phi$ the azimuthal angle of the total angular momentum. 
The intrinsic states for ChP violation system and their transformation under different combinations of $\mathcal{P}$, $\mathcal{T}$ and $\mathcal{R}$ operators are shown in Fig. \ref{fig1} (e). 
The symmetry restoration in the laboratory frame gives rise to four nearly degenerate rotational bands, i.e., ChP quartet bands, as shown in Fig. \ref{fig2} (e). 

The nuclear ChP violation is investigated with a reflection-asymmetric triaxial PRM in Ref. \cite{WangYY2020SB}, in which a new
symmetry for an ideal ChP violation system is found and the corresponding selection rules of the electromagnetic
transitions are derived. 
By taking a two-$j$ shell $h_{11/2}$ and $d_{5/2}$ with typical energy spacing for $A = 130$ nuclei, the fingerprints for the ChP violation including the nearly degenerate quartet bands and the selection rules of the electromagnetic transitions are discussed, it would be interesting to search for ChP quartet bands experimentally.

\subsection{Chiral Rotation Induced by Pairing Correlations}\label{pairing}

In the nuclear chiral rotation, the critical frequency is an important concept, which marks the onset of chiral rotation.
The chiral critical frequency has been investigated by the 3D cranking models based on a Woods-Saxon potential combined with the shell correction method \cite{Dimitrov2000PRL}, and relativistic \cite{ZhaoPW2017PLB,PengJ2020PLB} and  nonrelativistic \cite{Olbratowski2004PRL,Olbratowski2006PRC} density functional theories. 
However, in the microscopic and self-consistent density functional calculations, the pairing correlations are neglected.

In Ref. \cite{WangYP2023PLB}, based on the 3D cranking CDFT, a shell-model-like approach (SLAP) with exact particle number conservation is implemented to take into account the pairing correlations and the blocking effects exactly  and applied for the chiral doublet bands in $^{135}$Nd \cite{WangYP2023PLB}.  The data available, including the $I-\omega$ relation, as well as the electromagnetic transition probabilities $B(M1)$ and $B(E2)$, are well reproduced. It is found that the superfluidity can reduce the critical frequency and make the chiral rotation easier. 

\begin{figure*}[!]
  \centering
  \includegraphics[width=0.6\linewidth]{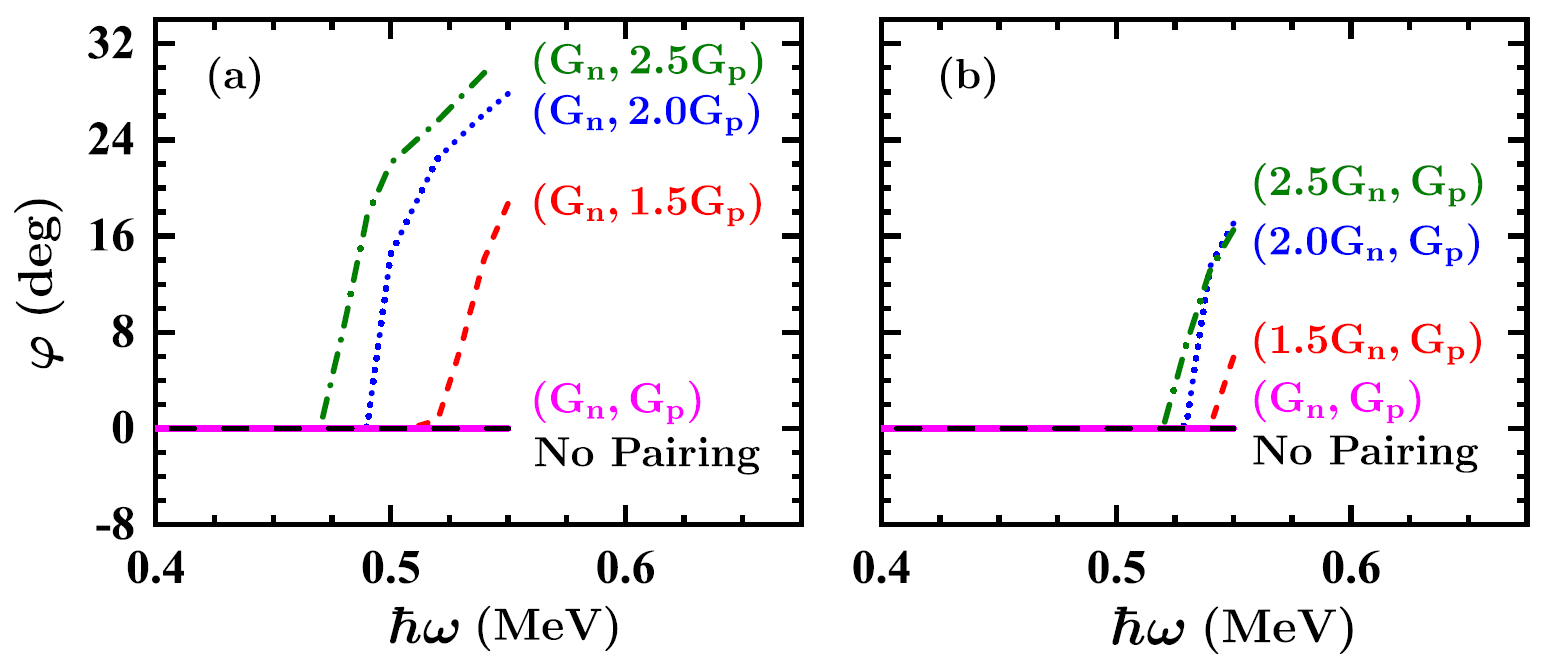}
  \caption{Evolution of the azimuthal angle for the total angular momentum as functions of the rotational 
           frequency $\omega$ in the cases without pairing or with different pairing strengths for protons (left panel) and neutrons (right panel). Reprinted figure from \cite{WangYP2023PLB}.}
  \label{fig8}
\end{figure*}

As shown in Fig. \ref{fig8}, without pairing, the azimuthal angle is always zero. 
In the left panel, when the proton pairing strength is enhanced by 50\%, 100\% or 150\%, the azimuthal angle respectively becomes nonzero at 0.52, 0.49 or 0.47 MeV, indicating the occurrence of the chiral rotation. 
Similarly, in the right panel, for the enhanced neutron pairing, the azimuthal angle respectively becomes nonzero at 0.54, 0.53 and 0.52 MeV.
These results suggest that the pairing correlations could induce the earlier appearance of the chiral rotation. 
 
\subsection{Chiral Dynamics}\label{TDCDFT}

The aforementioned studies on the nuclear chirality are based on static approaches, and the time-dependent (TD) CDFT is a dynamical extension of CDFT, which can serve as a powerful tool for simulating the dynamics of the nuclear chirality.
However, to achieve this goal, the solution of the CDFT in 3D lattice space is required, which is a challenging task.
Fortunately, after overcoming the longstanding problems of the variational collapse \cite{Hagino2010PRC} and Fermion doubling \cite{Tanimura2015PTEP}, the solution of CDFT in 3D lattice space has been realized \cite{Tanimura2015PTEP,RenZX2017PRC,RenZX2019SCP,RenZX2020NPA,LiB2020PRC,XuFF2024PRC}.

In Ref. \cite{RenZX2022PRC}, the dynamics of chiral nuclei is investigated with the TDCDFT in 3D lattice space in a microscopic and self-consistent
way. 
The experimental energies of the two pairs of the chiral doublet bands in $^{135}$Nd are well reproduced without
any adjustable parameters beyond the well-defined density functional. 
A novel mechanism, i.e., chiral precession, is revealed from the microscopic dynamics of the total angular momentum in the body-fixed frame, whose
harmonicity is associated with a transition from planar into aplanar rotations with the increasing spin. 

\begin{figure*}[!]
  \centering
  \includegraphics[width=0.6\linewidth]{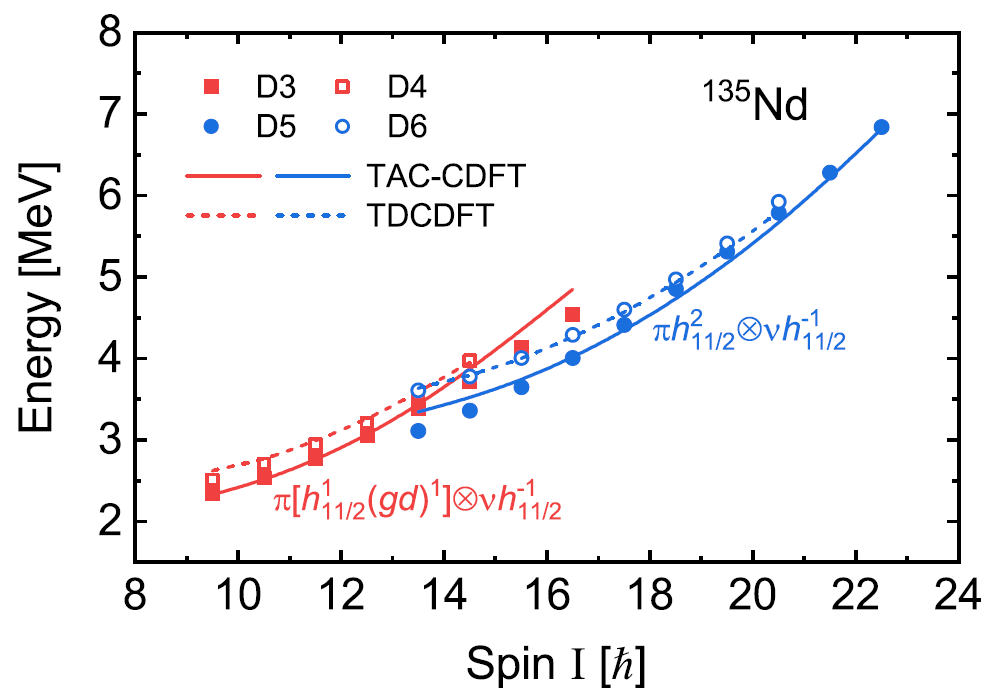}
  \caption{Calculated excitation energies (solid and dashed lines) for chiral doublet bands in $^{135}$Nd built on the
           configurations $\pi [h_{11/2}^1(gd)^1]\otimes \nu h_{11/2}^{-1}$ and $\pi h_{11/2}^2\otimes \nu h_{11/2}^{-1}$ in comparison with data (solid and open symbols) \cite{LvBF2019PRC}. The solid lines represent the TAC-CDFT results, and the dashed ones represent the results including the chiral excitation energies obtained with the TDCDFT calculations. Reprinted figure from \cite{RenZX2022PRC}.}
  \label{fig9}
\end{figure*}

As shown in Fig. \ref{fig9}, the calculated energies for the two pairs of chiral doublet bands in $^{135}$Nd, which are built, respectively, on the configurations $\pi[h_{11/2}^2(gd)^1]\otimes \nu h_{11/2}^{-1}$ and $\pi h_{11/2}^2\otimes \nu h_{11/2}^{-1}$, are depicted in comparison with the data \cite{LvBF2019PRC}. The solid lines represent the TAC-CDFT results, and the dashed ones represent the results including the chiral excitation energies obtained with the TDCDFT. 
It can be seen that the experimental
energies are well reproduced. This provides the first fully microscopic and self-consistent description for the chiral
doublet bands in the framework of the density functional theory.

\section{Summary and Perspective}
\label{sec8}

Since the concept of the nuclear chirality was proposed in 1997, the study of the nuclear chirality has been and will remain as one of the hot topics in nuclear physics. 
Over the past two and a half decades, many efforts have been devoted to understand chiral symmetry and its spontaneous breaking in atomic nuclei. 

Both experimental and theoretical efforts are still needed in the study of the nuclear chirality. 
From the experimental side, the simultaneous breaking of chirality and other symmetries would be interesting, for example, the search of ChP quartet bands. 
Finding more observables to provide further confirmation for chirality is also important. 
In 2018, the $g$ factor, which can give important information on the chiral geometry, has been measured for the bandhead state of the chiral doublet bands in $^{128}$Cs \cite{Grodner2018PRL}. 
The measurements of the $g$ factors for excited states and for more nuclei are looking forward.
Exploring new phenomenon is also necessary, such as the chiral wobbler in $^{74}$Br, with the
second and third lowest bands respectively suggested as the chiral partner and one-phonon wobbling
excitation built on the yrast band \cite{GuoRJ2024PRL}. 
From the theoretical side, the chirality in triaxial well-deformed nuclei has been extensively studied. In the future, it will be very interesting to further study the chirality in soft triaxial nuclei. 

\section{Acknowledgments}

We thank X. K. Du, T. X. Huang, F. Y. He, X. F. Jiang, T. Qu, F. F. Xu, Y. L. Yang and S. Q. Zhang for the careful reading and valuable suggestions.
This work was partly supported by the National Natural Science Foundation of China (Grants No. 11935003, No. 12105004, No. 12141501), the High-performance Computing Platform of Peking University, and the State Key Laboratory of Nuclear Physics and Technology, Peking University.

\end{document}